\documentclass[12pt]{article}
\usepackage{epsfig,amssymb}
\usepackage{epsfig}
\def\al{\alpha^{\prime}}
\def\a{& \hspace{-7pt}}

\def\bea{\begin{eqnarray}}
\def\eea{\end{eqnarray}}
\def\be{\begin{equation}}
\def\ee{\end{equation}}
\def\nn{\nonumber}
\def\a{& \hspace{-7pt}}

\def\Z{{\bf Z}}
\def\5{\overline 5}
\def\T6{${\bf T_6/Z_6^\prime\times Z_2^\prime }$ }

\begin{document}

\thispagestyle{empty}

\begin{center}
\hfill SISSA-64/2002/EP \\

\begin{center}

\vspace{1.7cm}

{\LARGE\bf On the quantum stability of Type IIB}
\vskip .3cm
{\LARGE\bf orbifolds and orientifolds with}
\vskip .3cm
{\LARGE\bf  Scherk-Schwarz SUSY breaking}

\end{center}

\vspace{1.4cm}

{\sc M. Borunda, M. Serone and M. Trapletti}\\

\vspace{1.2cm}

{\em ISAS-SISSA, Via Beirut 2-4, I-34013 Trieste, Italy} \\
{\em INFN, sez. di Trieste, Italy}
\vspace{.3cm}

\end{center}

\vspace{0.8cm}

\centerline{\bf Abstract}
\vspace{2 mm}
\begin{quote}\small
We study the quantum stability of Type IIB orbifold and orientifold string models in various dimensions,
including Melvin backgrounds, where supersymmetry (SUSY) is broken {\it \`a la} Scherk-Schwarz (SS)
by twisting periodicity conditions along a circle of radius $R$.
In particular, we compute the $R$-dependence of the one-loop induced vacuum energy density
$\rho(R)$, or cosmological constant. \\
For SS twists different from ${\bf Z}_2$ we always find, for both orbifolds and orientifolds,
a monotonic $\rho(R)<0$, eventually driving the system to a tachyonic instability.
For ${\bf Z}_2$ twists, orientifold models can have a different behavior, leading either
to a runaway decompactification limit or to a negative minimum at a finite value $R_0$.
The last possibility is obtained for a 4D chiral orientifold model where a more accurate but yet
preliminary analysis seems to indicate that $R_0\rightarrow \infty$ or towards the tachyonic instability,
as the dependence on the other geometric moduli is included.

\end{quote}

\vfill

\newpage
\setcounter{equation}{0}

\section{Introduction}

Supersymmetry (SUSY) is one of the most promising ideas to solve the hierarchy problem, if broken
at a low scale $M_{SUSY}\sim {\rm TeV}$. It plays an even more important role in a quantum
theory of gravity, such as string theory, since it provides a classical
and quantum stabilization mechanism for string/M-theory vacua.
The construction of absolutely stable string models, realistic or not, where SUSY is broken by means of any
symmetry breaking mechanism is a fundamental and outstanding problem in string theory.
In order to try to shed some light into this crucial question, it is essential to analyze
the form and nature of instabilities that affect the known non-SUSY string models.

One of the most interesting mechanisms of SUSY breaking and essentially
the unique available for oriented closed strings at the string level, 
is the Scherk-Schwarz (SS) mechanism \cite{SS}
in which SUSY is broken by twisting the periodicity conditions along some compact dimensions.
This mechanism allows to build classically stable models without tachyons
if the compactification radius along which the SS twist is performed
is chosen to be large enough. String models with SS SUSY breaking mechanism have received some
attention in both heterotic \cite{rohm,Ito,ssstringa,large,SeSc} and open strings
\cite{ads1,adds,sst}. On the contrary, not much analysis of the quantum stability of these vacua
have been performed, showing that the models evolve either to smaller radii reaching eventually
the tachyonic regime \cite{rohm}, or they recover unbroken SUSY by a decompactification limit
\cite{BD}\footnote{See \cite{Fabi} for a similar analysis performed in an M-theory context and
\cite{Ginsp} for a nice general analysis in non-SUSY heterotic models. See also \cite{ABG} 
for an analysis of the stability of a certain class of non-SUSY 6D orientifolds.}.

In this paper, along the lines of \cite{rohm,BD}, we compute the one-loop induced vacuum
energy density (cosmological constant) as a function of the radius of the twisted direction, for a certain
class of Type IIB orbifold and orientifold string models with SS SUSY breaking.
This is obtained by computing the one-loop partition function on the relevant world-sheet
surfaces: the torus for IIB orbifolds, and the torus, Klein bottle, annulus
and M\"obius strip for IIB orientifolds.

The first model we consider is a simple 9-dimensional (9D) 
$\Z_2$ IIB orientifold, whose vacuum energy density has
already been considered in \cite{BD}. In agreement with \cite{BD}, we show how the
cosmological constant of this model crucially depends on the choice made for the
$\Z_2$ Chan-Paton twist matrix $\gamma_g$. Depending on $\gamma_g$, the model evolves either
towards the tachyonic regime or to the decompactification limit (see figure 1).

Orbifolds and orientifolds on twisted
Asymptotically Locally Euclidean (ALE) spaces of the form $(\mathbb C\times S^1)/\Z_N$,
where ${\bf Z}_N$ ($N$ odd) acts simultaneously as a rotation on $\mathbb C$ and
a translation on $S^1$, are studied next. Such spaces upon reduction on $S^1$ give rise
to Melvin backgrounds \cite{melvin}, and have recently received renewed interest.
We show that for any $N$ odd, with or without open strings, the vacuum energy density
$\rho^N(R)$ is a negative monotonic function, reaching zero as $R\rightarrow\infty$.
In the orbifold models we find that $\rho^N(R)>\rho^M(R)$ for $N<M$, for any value of $R$, 
and for the lower $N,M=3,5,7,9$ considered (see figure 2). 
In the orientifold models, with a suitable choice of twist matrices, we find the same
behavior (see figure 3). In both cases, then, it is resonable to assume that $\rho^N(R)>\rho^M(R)$ 
for $N<M$, $\forall N,M$.
The perturbative quantum fate of these models is then clear. For any initial
value of $R$ and $N$, $R$ will shrink until $R=R_T$, the critical radius where some twisted string
mode becomes tachyonic. The dynamics is now governed by the classical tachyonic instability and,
according to the analysis of \cite{aps,dghm,suyama}, the model will undergo a phase transition towards
another $(\mathbb C\times S^1)/\Z_N^\prime$ model, with $N^\prime < N$, with $R$ increasing
along the transition \cite{suyama}.
The $\Z_{N^\prime}$ model will also undergo a phase transition and so on, until the
twist vanishes and a flat SUSY space-time is recovered. Quite interestingly,
we get exactly the same energy density $\rho^N(R)$ for non-compact $(\mathbb C\times S^1)/\Z_N$
or compact $(T^2\times S^1)/\Z_N$ models\footnote{In the compact case, $N$ has to
be odd and has to preserve the lattice structure of the torus.}, although the fate of the
two twisted directions seem to crucially depend on whether they are compact or not
\cite{aps,vafa}.

The last model we consider is a 4D chiral orientifold model, discussed in \cite{adds,sst}.
This is based on a $T^6/(\Z_6^\prime\times \Z_2^\prime)$ orbifold, where $\Z_6^\prime$ is the usual
rotation defining a SUSY orientifold \cite{iba} and $\Z_2^\prime = \sigma (-)^F$, with
$\sigma$ an order two translation along one of the radii ($R$) of a torus and $F$
the space-time fermion number operator. The model has $D9$-branes, as well as $D5$ and
$\bar D5$-branes. We first study $\rho$ as a function of the single modulus $R$,
fixing the remaining 5 other compactified directions. For a proper choice of the
Chan-Paton twist matrix associated to the $\Z_2^\prime$ element in the $D9$ sector,
we find a minimum of $\rho(R)$ at a finite value $R_0$, where $\rho(R_0)<0$,
providing in this way an interesting non-trivial stabilization mechanism for the SS direction 
$R$ (see figure 4).
In order to decide whether this minimum is actually an absolute minimum or not, we
also study the dependence of $R$ on some other compact directions, taken to be equal
for simplicity. We are not able to give a definite answer to whether $R_0$ is
an absolute minimum or not, but our analysis seems to suggest that for finite values of
the remaining compact directions, the minimum is always present, but it could
run-away to infinity or towards the tachyonic instability due to the dynamics along
the other directions (see figure 5).

An interesting by-product of our studies is provided by understanding which
string states contribute to $\rho(R)$. Generalizing the well-known technique of unfolding
the fundamental domain of the torus \cite{Tan}, we show that the closed string contribution
to $\rho(R)$, in both orbifold and orientifolds, can be analytically computed and
is given by untwisted closed strings only, where no winding modes along the SS direction appear.
As far as the $R$ dependence is concerned, the whole one-loop string partition function
looks effectively like that of a purely quantum field theory.
This provides a generalization of what is well-known to happen to strings 
at finite temperature \cite{pol,Tan} and to $\Z_2$ SS twists \cite{Ghil}.
For open strings, it is important to distinguish between longitudinal and transverse SS
breaking, depending on whether open strings can propagate or not along the SS direction \cite{ads1}.

The paper is organized as follows. In section 2 the general form of $\rho(R)$ for both closed
and open strings is shown, for both longitudinal and transverse SS breaking.
The 9D IIB orientifold is discussed in section 3 and in section 4 the class of orbifold/orientifold
models on twisted ALE spaces are considered. Section 5 is devoted to the 4D chiral orientifold.
Finally, some conclusions are given in section 6. In an appendix, we collect some details
on the technique of unfolding the torus fundamental domain to provide an analytic integration
of the torus one-loop partition function.

\section{General form of the vacuum energy}

The vacuum energy at one-loop level in any quantum field theory depends
simply on the mass spectrum of the theory
and its degeneracy. In $D$ space-time dimensions, in a Schwinger proper time parametrization,
it can be written as
\be
 E_D  =  - V_{D-1} \int_0^\infty \frac{dt}{2t} \sum_i d_i (-)^{2J_i}
\int \frac{d^Dk}{(2\pi )^D} e^{-t(k^2+m_i^2)},
\label{E0-FT}
\ee
with $J_i$ and $m_i$ the spin and mass of the state $i$, $d_i$ its degeneracy, $k$ its euclidean
momentum and $V_{D-1}$ the volume of the $D-1$ spatial dimensions.
More succinctly,
\be
E_D  =  - \int_0^\infty \frac{dt}{2t} \, {\rm Tr}_{\cal H} \,e^{-t H},
\label{partf}
\ee
where ${\cal H}$ is the full Hilbert space of our system (including $V_{D-1}$) and $H=k^2+m^2$.
We are interested in the explicit form of (\ref{E0-FT}) for orbifold and orientifold-derived
theories, with the usual tower of massive string states, where supersymmetry
is broken by a SS mechanism. We focus our attention to the case in which the twisted periodicity
conditions defining the SS breaking are taken only along one direction, a circle of radius $R$,
henceforth denoted SS direction. As we will see shortly, similar considerations apply also
when the SS direction is an $S^1/\Z_2$ orbifold, such as in the chiral 4D model considered
in section 5. More general configurations with two (or more) SS directions are possible
and have been explicitly constructed \cite{SeSc}, but will not be considered here.

It is convenient to treat separately the contribution to the vacuum energy of states propagating
or not along the SS direction. We denote the symmetry breaking in the two cases respectively as
longitudinal and transverse SS breaking.

\subsection{Longitudinal SS breaking}

This is the situation that will concern us mostly, since it applies to all closed strings
and to open strings on $D9$-branes. In both cases, the vacuum energy can be written as in
(\ref{E0-FT}). For open strings this is clear, $t$ being the modulus of the annulus and
$d_i$ the string degeneracy of the state. For closed strings the situation is more complicated
because the $PSL(2,{\mathbb Z})$ modular invariance of the torus
restricts the integration over the modulus $\tau$ to the fundamental domain. Nevertheless,
generalizing standard techniques (see {\it e.g.} \cite{Tan,MS} and the appendix) one can
unfold the fundamental domain to the strip and rewrite the whole closed string
contribution, including the Klein bottle term, in the form (\ref{E0-FT}), where $d_i$
is now the string degeneracy of the state, with the level-matching conditions
imposed by means of the $\tau_1$ integration. Quite interestingly, in this way only
untwisted  closed string states will explicitly appear in the computation.
Similarly, winding modes along the SS direction will not be present,
so that as far as the $R$ dependence is concerned, the whole amplitude
looks effectively like that of a {\it purely} quantum field theory with an infinite number of states.

It is useful to distinguish closed and open string contributions to the vacuum energy.
Recall that in string theory $H^{(closed)} = L_0+\bar L_0 = \al p^2/2 + (N+\bar N)/\al $ and
$H^{(open)} = L_0 = \al p^2 + N/\al $. Therefore, rescaling
$t\rightarrow \pi \al t$ and $t\rightarrow 2\pi \al t$ respectively in the
two cases, one gets
\bea
 \rho_D^{(C)}  & = &   -\int_0^\infty \frac{dt}{2\,t^{\frac{D+2}{2}}}
\, \sum_i d_i^{(C)} (-)^{2J_i}\,  e^{-\pi t\al m_i^2}, \nn \\
 \rho_D^{(O)}  & = &   -2^{-D/2}\int_0^\infty \frac{dt}{2\,t^{\frac{D+2}{2}}}
\, \sum_i d_i^{(O)} (-)^{2J_i}\,  e^{-2\pi t\al m_i^2},
\label{E0-FT2}
\eea
where $(C)$ and $(O)$ stand respectively for closed and open, and we have defined the energy densities
\be
\rho_D \equiv \frac{E_D(4\pi^2 \al)^{\frac D2}}{V_{D-1}},
\label{rhod}
\ee
with $V_{D-1}$ the $(D-1)$-dimensional spatial volume of the non-compact dimensions.
The generic mass of a given closed and open string Kaluza-Klein (KK) state at level $n$ 
along the SS direction is respectively
\bea
m^{2}_{i,n} = \frac{2 (N+\bar N)}{\al} + \ldots + \frac{(n+q)^2}{R^2}\equiv
m^{2}_{i}+ \frac{(n+q)^2}{R^2} \;,\nn \\
m^{2}_{i,n} = \frac{N}{\al} +\ldots  + \frac{[n+q(G)]^2}{R^2}\equiv
m^{2}_{i}+ \frac{[n+q(G)]^2}{R^2} \;,
\label{m2i}
\eea
where $q$ is the twisted charge given by the SS breaking, $N$ and $\bar N$ are the string oscillator
numbers and the dots stand for the KK and winding mode
contributions along the other compact directions. The index $i$ in (\ref{E0-FT2}) 
includes thus a sum over $N$, $\bar N$,
KK and winding modes over all the compact directions (but the SS direction)
of states of given charge $q$ and then a sum over all possible twists $q$.
Since the SS breaking can (and must) be implemented in the gauge sector as well, for open strings
$q$ depends also on the gauge degrees of freedom, $q=q(G)$.
The $m_i^{2}$ mass terms are typically functions of the geometric moduli of the
compactification, except $R$, but for simplicity of notation this dependence will be left
implicit in the following.
Eq.(\ref{E0-FT2}) can be rewritten as
\bea
 \rho_D^{(C)}  & = &   - \sum_{n,i} \sum_{F=0,1}
 \int_0^\infty \frac{dt}{2\,t^{\frac{D+2}{2}}}
\, d_i(q,F)\,  e^{i\pi F} \,  e^{-\pi t \al \big( m_i^2 +\frac{(n+q)^2}{R^2}\big)}, \nn \\
 \rho_D^{(O)}  & = & -2^{-D/2} \sum_{n,i,G} \sum_{F=0,1}
 \int_0^\infty \frac{dt}{2\,t^{\frac{D+2}{2}}}  d_i(q,F) \,e^{i\pi F}
 e^{-2 \pi t \al\big(m_i^2 +\frac{[n+q(G)]^2}{R^2}\big)},
\label{E0-FT3}
\eea
where $F$ is the space-time fermion number operator, $G$ denotes a sum over the gauge indices and
$d_i(q,F)$ are the string degeneracy factors, in general depending on $q$ and $F$,
that include also the degeneracy arising from the expansion of the modular functions.
By a Poisson resummation on the index $n$ and some algebra, it is not difficult
to explicitly compute $\rho_D^{(C)}$ and $\rho_D^{(O)}$. It is convenient to
separate the $m_i^2=0$ contribution, denoted by $\rho_{D,0}^{(C,O)}$,
from the remaining ones. One gets for both closed and open strings:
\be
\rho_{D,0}^{(C,O)}  =    -\bigg(\frac{\sqrt{\al}}{R}\bigg)^{D} \,
\frac{\Gamma(\frac{D+1}2)}{\pi^{\frac{D+1}2}} \sum_{q;F=0,1}
e^{i\pi F } d_0(q,F) \frac{{\rm Li}_{D+1}(e^{2i\pi q})+{\rm Li}_{D+1}(e^{-2i\pi q})}2\,,
\label{rho0} \ee
where
\be
{\rm Li}_p (z) = \sum_{w=1}^\infty \frac{z^w}{w^p}
\label{li}
\ee
are the polylogarithm functions, the sum over the gauge degrees of freedom for open strings is implicit,
and we denoted by $d_0$ the degeneracy of states with $m_i^2=0$.
As far as the $m_i^2\neq 0$ contributions are concerned, we get
\be
\rho_{D,i}^{(C,O)} =    -2 \bigg(
\frac{\sqrt{\al}}{R}\bigg)^{\frac{D-1}2}\!(\al m_i^2)^{\frac{D+1}4}\!\sum_{q;F=0,1}
e^{i\pi F } d_i(q,F) \!\sum_{w=1}^\infty
 \frac{\cos (2i\pi q w)}{w^{\frac{D+1}2}}
K_{\frac{D+1}2}(2\pi R w m_i), \label{rhoN} \ee
where $K_n$ are the modified Bessel functions and again the sum over gauge indices has been omitted.
The full vacuum energy $\rho_D$ is obtained by summing the total closed and open string
energy density contributions, $\rho_D=\rho_D^{(C)}+\rho_D^{(O)}$, where
\be
\rho_D^{(C,O)}= \rho_{D,0}^{(C,O)}+\sum_{i\neq 0} \rho_{D,i}^{(C,O)}
\label{rhoTOT}
\ee
and $i\neq 0$ indicates that states with $m_i^2=0$ should not be included in the sum.
Both $\rho_D^{(C)}$ and $\rho_D^{(O)}$ are finite, as expected by the non-local nature
of the SS breaking. Potential divergences should be local in space-time, but locally SUSY
is preserved and hence no divergences at all are present.
Indeed, in both eqs.(\ref{rho0}) and (\ref{rhoN}) there would be a potential
$R$-independent UV divergence arising from the $w=0$ term, where the index $w$, entering
in (\ref{li}), is obtained by a Poisson resummation on the index $n$ of eq.(\ref{E0-FT3}).
This term vanishes because at each mass level $i$ the total number of bosons,
summed over all possible twists $q$, equal the total number of fermions:
\be
\sum_q d_i(q,0) = \sum_q d_i(q,1)  \ \ \forall i\,.
\label{boseqfer}
\ee
String models with SS SUSY breaking generically have winding modes that become
tachyonic below certain values of $R$, where the vacuum energy diverges.
Though the $m_i^2$ mass terms defined in (\ref{m2i}) are always positive, this divergence
appears in (\ref{rhoN}) from the sum over all massive string states.
As is well known, the degeneracy of massive string states, for large masses,
has a leading exponential behavior $d_i\sim \exp(2\pi c \sqrt{\al}m)$,
with $c$ a given constant.
On the other hand, for large values of its argument,
the modified Bessel function $K_n(z)$ admits an asymptotic expansion whose
leading term is $\sim \exp(-z)$. Hence, we see that the infinite sum over $i$
in (\ref{rhoN}) converges only for $R> R_T=\sqrt{\al} c$. When the SS twist $q=F$,
one can easily recognize eqs.(\ref{rho0}) and (\ref{rhoN}) to be closely related to the
free energy of string/field theory-derived models, with $1/T=2\pi R$ (see {\it e.g.}
\cite{alva}) and $T_H=1/(2\pi R_T)$ being the Hagedorn temperature.

The general form of the vacuum energy for $m_i R>> 1$ is easily extracted.
We see from (\ref{rho0}) that the $m_i^2=0$ contributions is power-like in $R$,
whereas for large $R$ (\ref{rhoN}) is exponentially suppressed in $R$.
More precisely we get
\be
\rho_D\sim \frac{1}{R^D} \sum_q \big[ d_0(q,1)-d_0(q,0) \big] C_D(q) +
O\bigg(\frac{e^{-m_i R}}{R^{\frac D2}}\bigg),
\ee
where $d_0(q,1)$ and $d_0(q,0)$ are the total (closed + open) number of fermionic and bosonic massless
states in $D+1$ dimensions (before the SS compactification) with charge $q$
and $C_D(q)$ are certain functions easily obtained from (\ref{rho0}).
For $\Z_2$ or $\Z_3$ twists, in which we get two independent twists $q=0,1/2$ or $q=0,1/3$,
the constraint (\ref{boseqfer}) implies that
for large $R$ the vacuum energy is dominated by the difference between the total number of
fermionic and bosonic $D$-dimensional, rather than $D+1$-dimensional, massless states 
$d_0(0,1)$ and $d_0(0,0)$ \cite{Ito};
\be
\rho_D\sim \frac{d_0(0,1)-d_0(0,0)}{R^D} + O\bigg(\frac{e^{-m_i R}}{R^{\frac D2}}\bigg)\,.
\ee
In these cases, an exponentially small one-loop cosmological constant requires
$d_0(0,0)=d_0(0,1)$ \cite{Ito,large}.
However, for more general twists, we notice that the leading power-like behavior
can be vanishing in a non-trivial way, thanks to a compensation between
bosonic and fermionic contributions with different twists, even if $d_0(0,0)\neq d_0(0,1)$.
It would be quite interesting to fully exploit this observation and see if there exist
string models with a spectrum satisfying this property.

All the above considerations are easily generalized to the case in which the SS direction
is an $S^1/\Z_2$ orbifold. Bulk states propagating along the orbifold are now classified
according to their $\Z_2$ parities. The massive spectra of $\Z_2$-even and $\Z_2$-odd
states differ only by the presence or not of a zero mode along the orbifold.
Both contributions can be summed together in the form (\ref{E0-FT3}), where the KK level
runs over all integers. Possible left-over $n=0$ terms in the process of recombining
$\Z_2$-even and $\Z_2$-odd contributions must vanish, since SUSY is broken by
the compactification and they do not depend on $R$.
Equation (\ref{E0-FT3}) and all the analysis that follows is then still valid for the
$S^1/\Z_2$ orbifold.
The remaining compact directions can be instead arbitrary, as far as SUSY is broken only
by the twist on $R$. Their structure will affect the explicit form of $m_i^2$ as well as
the degeneracy factors $d_i(q,F)$.

\subsection{Transverse SS breaking}

States that do not propagate along the SS direction do not have a KK decomposition
along that direction and hence their contribution to the vacuum energy requires a separate
analysis. In string theory, states of this kind can arise either as twisted closed strings located
at fixed points orthogonal to the SS direction or as open strings on $D$-branes
transverse to the SS direction. In our class of models, the first kind of states appears always
with unbroken tree-level SUSY and hence will never contribute to the one-loop vacuum energy.
The same applies also to open strings on $D$-branes transverse to the SS direction, that
present unbroken SUSY at the classical level. The only exception arises for open strings stretched between
$D5$-branes/$O5$-planes and $\bar D5$-branes/$\bar O5$-planes, 
where SUSY is broken at tree level, for the 4D model
discussed in section 5. In this case, the one-loop open string amplitude is more conveniently
expressed as a tree-level exchange of closed string states, propagating from one object to the other.
This contribution can be thus summarized as follows:
\be
\rho_D \sim -V_n \sum_i \sum_{\hat w} Q^{(i)}\bar Q^{(i)} {\cal G}_d^{(i)}
[\Delta x(\hat w)]\,,
\label{Rhotran}
\ee
where ${\cal G}_d^{(i)}$ is the $d$-dimensional propagator of a particle of mass $m_i$ and spin $J_i$
and $V_n$ is the volume of the compact longitudinal directions 
along which the states can propagate.
$\Delta x(\hat w)$ is the transverse $D$-brane/$O$-plane-${\bar D}$-brane/${\bar O}$-plane
distance modulo windings, since a given closed string
state can wind $w$ times along all transverse directions before ending on a $D$-brane/$O$-plane, and
$Q^{(i)}=Q_D^{(i)}/Q_O^{(i)}$ is the $D$-brane/$O$-plane charge for each state $i$.
The same applies to the anti-brane and anti-plane charges $\bar O^{(i)}$.
As in last subsection, the sum over $i$ runs over the string and winding modes
along the longitudinal directions, whereas the sum over $\hat w$ runs over all the possible
windings along the transverse $d$ dimensions.
One should recall that the subscripts $d$ and $n$ in (\ref{Rhotran}) represent 
the number of space-time directions in which
closed strings propagate and thus it can be different for untwisted and various twisted closed
string states.
As in the last subsection, the massless $i=0$ contributions in (\ref{Rhotran}) are power-like in $R$,
whereas massive ones give an infinite sum over modified Bessel functions $K_n$ and
thus are exponentially suppresed in $R$, for large $R$.
The sign of the vacuum energy contribution is now given by the brane/plane charge and is 
always negative. This is intuitively clear, since objects with opposite charges feel
an attractive potential between each other.
The divergence due to the open string tachyon arising below a given radius is 
determined by looking at the asymptotic form of the modified Bessel functions
and at their degeneracies for large masses. 

\section{A nine-dimensional model}

We compute in this section the energy density $\rho_9$ of a simple 9D
Type IIB orientifold compactified on $S^1/(\Omega\times {\bf Z}_2)$, where
$\Omega$ is the world-sheet parity operator and ${\bf Z}_2$ is generated by
$g$, the product of a $\pi R$ translation $\sigma$
along the circle and $(-)^F$, with $F$ the space-time fermion number operator;
$g=\sigma (-)^F$. Such a computation has already been done in \cite{BD};
we review it in the following as a simple example of the kind of computation
we are going to perform in the next sections.
We are interested in the radius dependence of $\rho_9=\rho_9(R)$;
for simplicity we do not include continuous Wilson lines but discuss the dependence
of $\rho_9(R)$ on the twist matrices $\gamma_g$ that embed the ${\bf Z}_2$ group
in the Chan-Paton degrees of freedom. They can alternatively be considered as ${\bf Z}_2$
discrete Wilson lines.

The construction of the model is straightforward and will not be presented in detail
(see {\it e.g.} \cite{BD,ads1}). The only massless tadpoles are those for the dilaton, graviton (NSNS)
and for the untwisted 10-form (RR), whose cancellation requires the presence of 32 $D9$-branes.
All neutral fermions are anti-periodic along the circle and thus massive, with
a mass $\sim 1/R$. In the twisted sector we get a tower of real would-be tachyons
starting from $\al m^2 = R^2/(4\al) -2$. The twist matrix $\gamma_g$ is arbitrary
and must only satisfy the group algebra $\gamma_g^2=\pm I$.
The gauge group is $SO(n)\times SO(32-n)$, with massless fermions in the
bifundamentals $({\bf n},{\bf 32-n})$, for $\gamma_g={\rm diag} (I_n, -I_{32-n})$.
If $\gamma_g$ is taken traceless, the twisted tadpoles associated to the
above would-be tachyons cancel as well. From this perspective, such a choice is
different from the others.
If $\gamma_g$ is chosen antisymmetric, one gets a $U(16)$ group, or subgroups thereof,
with massless fermions in antisymmetric representations\footnote{In this model, local tadpole cancellation
for both untwisted NSNS and RR tadpoles cannot be achieved. By choosing respectively $\gamma_g$
symmetric or antisymmetric one can cancel respectively the massive NSNS or RR tadpoles.}.
We focus in the following on symmetric twist matrices for definiteness, but
since the energy density $\rho_9(R)$ depends on $\gamma_g$ only through its trace,
the analysis applies equally well for other more general choices.

The full energy density of the model is obtained by summing the closed string
contribution $\rho_9^{(C)}$ (torus+Klein bottle) and the open one, $\rho_9^{(O)}$
(annulus+M\"{o}bius strip).

\subsection{Closed string contribution}

Since the Klein bottle amplitude vanishes identically (the $\Omega$-projection acts
in a supersymmetric manner on the closed string spectrum), the whole
contribution is given by the torus amplitude:
\be
\rho_9^{(C)}=-\int_{\cal F}\!
\frac{d^2\tau}{2\tau_2^{11/2}}\frac{1}{2|\eta|^{24}}\sum_{n,w\in \mathbb Z}
\bigg\{|\theta_2^4|^2\Lambda_{n,w}(-)^n
+ |\theta_4^4|^2\Lambda_{n,w+\frac 12} +
|\theta_3^4|^2\Lambda_{n,w+\frac 12}(-)^n \bigg\}\,,
\label{rhod9c}
\ee
where
\be
\Lambda_{n,w}=q^{\frac{\alpha^{\prime}}{4}\big(\frac{n}{R}+\frac{wR}{\alpha^{\prime}}\big)^2}
            \overline{q}^{\frac{\alpha^{\prime}}{4}\big(\frac{n}{R}-\frac{wR}{\al}\big)^2}\,,
\ee
and $\theta_i=\theta_i(0|\tau)$ are the usual theta functions, defined as in \cite{iba}.
In eq.(\ref{rhod9c}) and throughout all the paper, we often omit the
modular dependence of $\theta_i$ on $\tau$ and always leave implicit its vanishing argument $z=0$.
Using the unfolding technique (see the appendix), (\ref{rhod9c}) can be written as
\be
\rho_9^{(C)} = -\frac{R}{\sqrt{\al}} \int_0^\infty
\frac{d\tau_2}{2\tau_2^6}
\sum_{N\in \mathbb N} d^2_N\, e^{-4\pi\tau_2 N}
\sum_{p=\in \mathbb Z} \bigg[\frac{1-(-)^p}{2} \bigg]
e^{-p^2\textstyle{\frac{\pi R^2}{4\tau_2\al}}}\,,
\label{lambdato}
\ee
where
\be
\frac 12\int_{-1/2}^{1/2}d\tau_1 {|\theta_2^4|^2\over|\eta|^{24}}
\equiv\sum_{N\in \mathbb N} d^2_N\, e^{-4\pi\tau_2 N}\,.
\label{expt1}
\ee
Eq.(\ref{lambdato}) can also be written by using a Poisson resummation as
\be
\rho_9^{(C)} = -\int_0^\infty
{d\tau_2\over {2\,\tau_2^{11/2}}} \sum_{N\in \mathbb N} d^2_N
e^{-4\pi \tau_2 N}\sum_{n\in \mathbb Z}
\bigg[ e^{-\pi \tau_2 \frac{\al n^2}{(R/2)^2}} -
e^{-\pi\tau_2\frac{\al (n+1/2)^2}{(R/2)^2}} \bigg]\,.
\label{CCstring}
\ee
Eq.(\ref{CCstring}) is precisely of the general form (\ref{E0-FT3}) with $D=9$, $q=F/2$, $i=N$,
$\al m_N^2=4N$, $d_N(0,0)=d_N(1/2,1)=d^2_N$, $d_N(0,1)=d_N(1/2,0)=0$,
and $R/2\rightarrow R$. The rescaling of the radius is standard and is due to the
identification $x\sim x+\pi R$ induced by the freely acting action.
We thus notice, as mentioned in the last section, that the full string theory contribution
to the cosmological constant (all KK, winding, untwisted and twisted string states) is
automatically encoded in a field theory contribution with only untwisted states,
whose KK modes are shifted by the SS mechanism (bosons/fermions periodic/anti-periodic
along the SS circle R) and a reduced massive string spectrum, where
only the diagonal $N=\bar N$ states contribute \cite{Ghil}.
The integration on $\tau_2$ in (\ref{CCstring}) can thus be read off directly from
(\ref{rho0}) for $N=0$ and (\ref{rhoN}) for $N\neq 0$.

\subsection{Open string contribution}

The annulus and M\"{o}bius strip contributions, in the absence of Wilson lines, are 
\bea
\rho_9^A = \a\a - \big({\rm Tr}\, \gamma_g\big)^2
\int_0^\infty\!\!\frac{dt}{4(2t)^{11/2}}
\frac{\theta_2^4}{\eta^{12}}(it) \sum_{m\in\mathbb Z} 
(-)^m e^{-2\pi tm^2\frac{\al}{R^2}}\,, \nn \\
\rho_9^M = \a\a + 32 \int_0^\infty\!\!\frac{dt}{4(2t)^{11/2}}
\frac{\theta_2^4}{\eta^{12}}(it-\frac{1}{2}) \sum_{m\in\mathbb Z} 
(-)^m  e^{-2\pi t m^2\frac{\al}{R^2}} .
\label{apertaIB}
\eea
\begin{center}
\begin{figure}[t]
\hspace{2.5cm}
\includegraphics[scale=0.55]{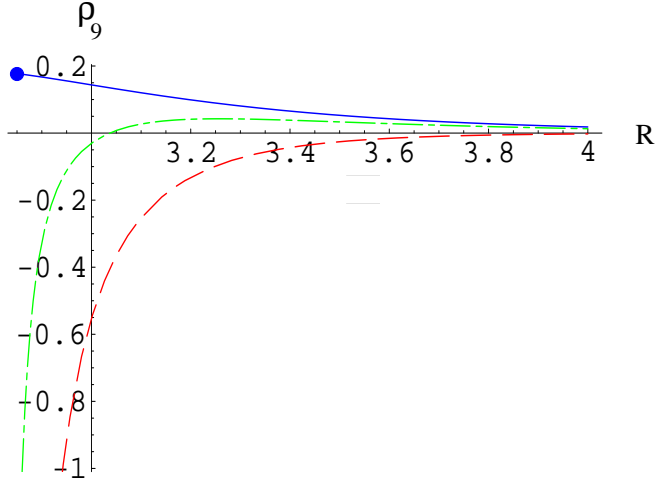}\vspace{-.3cm}
\caption{\small Behavior of the vacuum energy density for different choices of the twist matrix $\gamma_g$.
The upper line refers to the $SO(16)\times SO(16)$ group,
the intermediate line to $SO(17)\times SO(15)$ and the lowest one to $SO(18)\times SO(14)$.
$R$ is in units of $\al$. The bullet in the solid line represents the point where a tachyon appears.}
\label{Fig1}
\end{figure}
\end{center}\vspace{-1cm}
The integrations of (\ref{apertaIB}) are easily performed and one gets
\bea
\rho_{9,0}^A+\rho_{9,0}^M = \a\a \bigg[ 32 -({\rm Tr}\, \gamma_g)^2\bigg]
\bigg(\frac{\sqrt{\al}}R\bigg)^9 2^{\frac{15}2} \, d_0 \,
\frac{\Gamma(5)}{\pi^5}
\, \xi (10) ( 2-2^{-9})\,,  \label{lambda9da} \\
\rho_{9,N}^A+\rho_{9,N}^M = \a\a \bigg[ 32 (-)^N -({\rm Tr}\, \gamma_g)^2\bigg]
\bigg(\frac{\sqrt{\al}}R\bigg)^4 2^{\frac 72} \, d_N \,N^{\frac 52}
\sum_{n=1}^\infty \frac{1-(-)^n}{n^5} K_5\bigg[\pi R
n\sqrt{{N\over\alpha^\prime}}\bigg]\,, \nn
\eea
with $d_N$ as in (\ref{expt1}). It is not difficult to realize that these
expressions are of the general form (\ref{E0-FT3}) with $D=9$, $i=N$, $\al m_N^2=N$
and $q_{ij}(G) = F/2 + \hat q_i +\hat q_j $, where $i,j$ run over
the fundamental representation of $G=SO(32)$. The parameters $\hat q_i=0,1/2$ are related to the
eigenvalues of $\gamma_g$ by
\be
\gamma_g = {\rm diag}\, (e^{2i\pi \hat q_1}, \ldots,e^{2i\pi \hat q_{32}})\,.
\label{gammag}
\ee
As mentioned, the resulting 9D gauge group is $SO(n)\times SO(32-n)$ if
$\gamma_g={\rm diag} (I_n, -I_{32-n})$, with bosons and
fermions in antisymmetric, symmetric and bifundamental representations, depending
on the KK and string mass level $N$. Notice that the leading open string contribution to the
energy density is positive whenever $14\leq n\leq 18$, with a maximum for $n=16$.

The full energy density $\rho_9 = \rho_9^{(C)}+  \rho_9^{(O)} =  \rho_9^{(C)}+ \rho_9^{A}
+ \rho_9^{M}$ can then be numerically evaluated as a
function of $R$ by truncating the infinite sums
of modified Bessel functions.

The $R$-dependence of $\rho_9$ crucially depends on $\gamma_g$. For $n\leq 14$ or $n\geq 18$,
$\rho_9<0$ monotonically and leads the system to a tachyonic instability,
like in the early work of \cite{rohm}. For $15\leq n\leq 17$, $\rho_9$ can be positive
and a {\it maximum} close to $R_T$ is obtained for $n=15$ or $n=17$.

In figure (1) we show these different behaviors plotting $\rho_9(R)$
for $n=16$ (blue/solid line), $n=15 \sim n=17$
(green/dotted-dashed line) and $n=14 \sim n=18$ (red/dashed line).

\section{Strings on twisted ALE spaces}

An interesting class of models whose energy density can be studied
are orbifold or orientifold models on Asymptotically Locally Euclidean (ALE) spaces,
non-trivially fibered
along an $S^1$, of the form $({\mathbb C}\times S^1)/{\bf Z}_N$,
or their compact versions $(T^2\times S^1)/{\bf Z}_N$. The ${\bf Z}_N$ generator
is a product of a $4\pi /N$ rotation along ${\mathbb C}$ and of a translation
of $2\pi R/N$ along the circle. The rotation is taken to be of angle $4\pi /N$ so that
$g^N=1$ on spinors.
A non-trivial fibration
is necessary to implement the Scherk-Schwarz supersymmetry breaking and to lift the
mass of the would-be tachyons. Upon compactification on $S^1$, such spaces give
rise to a Melvin background \cite{melvin} (see \cite{melvinstring} and
\cite{Dmelvin} for strings and $D$-branes
on Melvin backgrounds).

Consider then Type IIB string theory
on $({\mathbb C}\times S^1)/{\bf Z}_N$ or
$(T^2\times S^1)/{\bf Z}_N$ (as we will see, our results
are the same for the non-compact or compact version). In order to keep our analysis
as simple as possible, we focus on $N$ odd.
Other values of $N$ would require the introduction of $O7$-planes and $D7$-branes,
in addition to $D9$-branes and the $O9$-plane, when considering orientifolds.
Moreover, for $N$ odd, would-be tachyons appear localized in space-time and an understanding
of the possible tachyonic instabilities can be obtained \cite{aps,dghm,suyama,vafa}.

Uncharged fermions in all these models are massive for $N \neq 3$,
with a mass $\sim 1/R$. In each $k$-twisted sector
we get a tower of complex would-be tachyons starting from
$\al m^2 = -4\frac{k}{N}+\left(\frac{k}{N}\right)^2\frac{R^2}{\al} $.


\subsection{Orbifold models}

The only relevant one-loop world-sheet surface in this case is the torus.
Its contribution to the vacuum energy density $\rho_7^{N,T}(R)$
can be written, using the unfolding technique discussed in the appendix
(see \cite{mich} for a more detailed discussion), as an integral
over the strip of the untwisted sector only
\be
\label{torozn}
\rho_7^{N,T} =- {1\over N}{R\over\sqrt{\al}}
\int_0^\infty\!\!\frac{d\tau_2}{2 \tau_2^5}\sum_{k=1}^{N-1}
\sum_{M\in\mathbb N}\left[ d_M(B)^k - d_M(F)^k \right] e^{-4\pi\tau_2 M}
\sum_{n\in\mathbb{Z}}
e^{-\frac{\pi R^2}{\alpha^\prime \tau_2}\left(n+\frac{k}{N}\right)^2},
\ee
where we have defined for later convenience the coefficients $d_M(B,F)^k$ as
\be
\sum_{M\in\mathbb N}\left[ d_M(B)^k - d_M(F)^k \right] e^{-4\pi\tau_2 M}
=\int\!\! \frac{d\tau_1}{4}
\Bigg|
2\sin ({2\pi k\over N})\frac{\sum_{\alpha,\beta}\eta_{\alpha\beta}
\theta{\alpha\brack \beta}^3\theta{\alpha\brack \beta+\frac{2k}{N}}}
{\eta^9\,\,\theta{\frac 12\brack \frac 12+\frac{2k}{N}} }
\Bigg|^2\,.
\label{defdm}
\ee
\begin{center}
\begin{figure}[t]
\hspace{2cm}
\includegraphics[scale=.65]{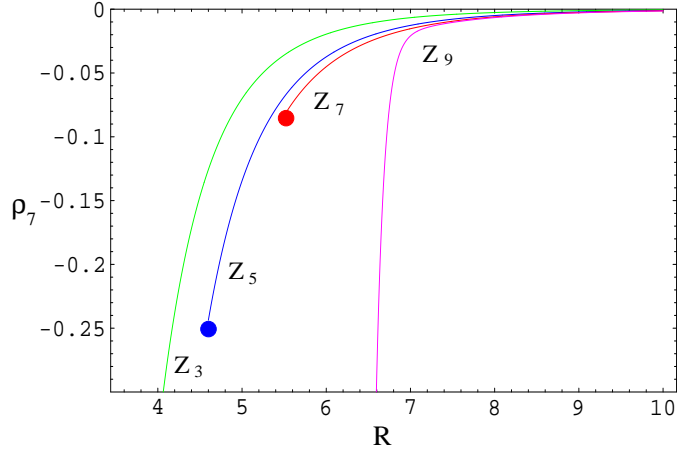}
\vspace{-.8cm}
\caption{\small $R$-dependence of the torus contribution to the cosmological constant
for $\Z_N$ orbifolds. $R$ is the radius of the
SS circle, in units of $\al$. The bullets represent the points where
tachyons appear.}
\label{ab}
\end{figure}
\end{center}

If, similarly to the 9D case of last section, the torus
contribution to the vacuum energy is exactly encoded in a field theory-like
contribution with only KK and untwisted states, (\ref{torozn})
should admit a rewriting such as (\ref{E0-FT3}), where the twist $q$ will be given by
the Lorentz $SO(2)$ charge of the states along the twisted directions, with the
correct degeneracies. It is useful to compute the latter in some detail for the massless
states of IIB string theory.

The above orbifold breaks the Lorentz group $SO(10)\rightarrow SO(8)\times SO(2)$.
A generic field $\Phi$ will have the following periodicity conditions along
$S^1$;\footnote{Notice
that due to the freely acting action, the radius $R$ entering in (\ref{phi}) is $N$ times
smaller than the $R$ appearing in eq.(\ref{torozn}).}
\be
\Phi (y+2\pi R) = e^{2i\pi \hat q \alpha}\Phi (y),
\label{phi}
\ee
where $\hat q$ is its charge under the $SO(2)\simeq U(1)$ internal Lorentz group and
$\alpha = 2/N$ is the twist induced by the $\Z_N$ action.
Upon reduction on $S^1$, a field with charge $\hat q$ will have a tower
of KK states with masses $M_n^{(q)} = (n + q)/R$, where $q=\hat q \alpha$.
Massless states are not present whenever $q \notin {\mathbb Z}$.

The $U(1)$ charges of the massless states of IIB string theory
are easily obtained. The Ramond-Ramond (RR) four, two and zero-forms have the following
$SO(8)\times U(1)$ decomposition:
\bea
{\bf 35} \a = \a {\bf 15}_0 \oplus {\bf 10}_1 \oplus {\bf 10}_{-1}, \nn \\
{\bf 28} \a = \a {\bf 15}_0 \oplus {\bf 6}_1 \oplus {\bf 6}_{-1} \oplus {\bf 1}_0 , \nn \\
{\bf 1} \a = \a {\bf 1}_0,
\eea
where the subscript denotes the $U(1)$ Lorentz charge $\hat q$.
One gets for the Neveu-Schwarz/Neveu-Schwarz (NSNS) graviton, $B$-field and dilaton:
\bea
{\bf 35} \a = \a {\bf 21}_0 \oplus {\bf 6}_1 \oplus {\bf 6}_{-1} \oplus
{\bf 1}_2 \oplus {\bf 1}_{-2} , \nn \\
{\bf 28} \a = \a {\bf 15}_0 \oplus {\bf 6}_1 \oplus {\bf 6}_{-1} \oplus {\bf 1}_0, \nn \\
{\bf 1} \a = \a {\bf 1}_0 .
\label{NSNS}
\eea
For generic $N\neq 2,4$ we thus get 70 bosonic massless states.
For fermions one has two copies of:
\bea
{\bf 56} \a = \a {\bf 24}_{1/2} \oplus {\bf 24}_{-1/2} \oplus {\bf 4}_{3/2}\oplus {\bf 4}_{-3/2} , \nn \\
{\bf 8} \a = \a {\bf 4}_{1/2} \oplus {\bf 4}_{-1/2}.
\label{NSR}
\eea
Notice that all fermions are always massive except for the case $N= 3$, where
we get 16 massless states from the decomposition of the two gravitinos.

We are now ready to explicitly show how these massless states and the corresponding twists
arise from (\ref{torozn}). We focus on the $N=3$ case, but
the analysis can be generalized to the other cases.
Reintroducing also the vanishing $k=0$ term in (\ref{torozn}) we have
($\tau_2\rightarrow t$ and $R\rightarrow NR$):
\bea
\label{torozn4}
\rho_7^T = -{R\over\sqrt{\al}}
\int_0^\infty{dt\over 2t^5} \sum_{M\in \mathbb N} \sum_{n\in\mathbb{Z}}
e^{-4\pi t M}\!\!\!\! \a\a \bigg\{\big[ d_M^0(B)-d_M^0(F) \big]
e^{-(3n)^2\frac{\pi R^2}{\al t}}+ \nn \\
\a\a \big[ d_M^1(B)-d_M^1(F) \big]
e^{-(3n+1)^2\frac{\pi R^2}{\al t}}+ \nn \\
\a\a \big[ d_M^2(B)-d_M^2(F) \big]
e^{-(3n+2)^2\frac{\pi R^2}{\al t}} \bigg\}.
\eea
Notice that the above sums can be rewritten as follows:
\bea
\sum_{n\in\mathbb{Z}} e^{-(3n)^2\frac{\pi  R^2}{\al t}} \a = \a
\sum_{n\in\mathbb{Z}}\frac{1+e^{\frac{2i\pi n}3}+e^{\frac{-2i\pi n}3}}3\,
e^{-n^2\frac{\pi R^2}{\al t}},  \\
\sum_{n\in\mathbb{Z}} \bigg[e^{-(3n+1)^2\frac{\pi R^2}{\al t}}+
e^{-(3n+2)^2\frac{\pi  R^2}{\al t}}\bigg] \a = \a
\sum_{n\in\mathbb{Z}}\frac{2-e^{\frac{2i\pi n}3}-e^{\frac{-2i\pi n}3}}3\,
e^{-n^2\frac{\pi  R^2}{\al t}}. \nn
\eea
Since $d_M(B,F)^1=d_M(B,F)^2\,$ $\forall M$, eq.(\ref{torozn}) can be rewritten,
by performing a Poisson resummation in $n$, as
\bea
\label{torozn3}
\rho_7^T=  - \int_0^\infty {dt \over 2t^{\frac{9}{2}}}\sum_{n\in\mathbb{Z}}
\sum_{M\in\mathbb N} \,e^{-4\pi t M} \Bigg\{ \a\a
\bigg[ d_M(0,0) - d_{M}(0,1) \bigg] e^{-\frac{\pi t \al n^2}{R^2}} \nn \\
+ \a\a \bigg[ d_M(\frac 13,0) - d_{M}(\frac 13,1) \bigg]
e^{-\frac{\pi t \al (n+1/3)^2}{R^2}}\Bigg\}\,,
\eea
where
\bea
d_M(0,0/1) \a = \a \frac 13\big[d_M(B/F)^0 + 2 d_M(B/F)^1 \big]\,, \nn \\
d_M(\frac 13,0/1) \a = \a \frac 23\big[ d_M(B/F)^0 - d_M(B/F)^1\big]
\eea
are the total number of bosonic and fermionic states at level $M$ with twist
0 and 1/3 respectively.
One can easily check that for $M=0$ the above coefficients precisely coincide
with the field theory results above.
Indeed, from (\ref{defdm}) one finds $d_0(B)^0=d_0(F)^0=128$, $d_0(B)^1=d_0(B)^2=41$,
$d_0(F)^1=d_0(F)^2=-40$ and hence $d_0(0,0)=70$, $d_0(0,1)=16$, $d_0(1/3,0) =58$,
$d_0(1/3,1)=112$,
as expected. Eq.(\ref{torozn3}) is hence precisely of the form (\ref{E0-FT3})
with $D=7$, $i=M$, $\al m^2_M=4M$ and $q=0,1/3$.

Interestingly, even for ${\bf Z}_3$ (and most likely for all ${\bf Z}_N$, at
least with $N$ odd) the {\it full} string theory
computation is encoded in a field theory computation where only
untwisted states with the uncompactified level matching conditions $M=\bar M$
enter.

The form of $\rho_7^N(R)$ for various $\Z_N$ orbifold models is reported in figure 2.
As anticipated in the introduction, we find that $\rho_7^N(R)>\rho_7^M(R)$ for $N<M$
and for all the values of $R$ for which both energy densities are well defined.

\subsection{Orientifold models}

Orientifold models are obtained as usual by modding out the above orbifolds
by the world-sheet parity operator $\Omega$.
As in the 9D model discussed in
last section, massless tadpoles are cancelled by introducing 32 $D9$-branes.
We do not include continuous Wilson lines but consider the dependence
of $\rho_7^N(R)$ on the twist matrices $\gamma_g$, whose only constraint comes from
the group algebra: $\gamma_g^N=\pm I$.
The resulting gauge group depends on the precise form of $\gamma_g$, although
the energy density is sensitive only on its trace.
In addition to the torus amplitude (\ref{torozn}), we have now to consider also the Klein bottle,
annulus and M\"obius strip surfaces.

The Klein bottle  contribution, $\rho_7^{N,K}$, is easily obtained and has a form
similar to (\ref{torozn}).
Defining
\bea
\sum_{M\in\mathbb N} [D_M(B)^{2k} -D_M(F)^{2k}] e^{-4\pi M\,t}=-2
\sin ({4\pi k\over N})
\frac{\sum_{\alpha,\beta}\eta_{\alpha\beta}\,
\theta{\alpha\brack\beta}^3\theta{\alpha\brack\beta+\frac{4k}{N}}}{
2\,\eta^9\,\,\theta{\frac 12\brack \frac 12+\frac{4k}{N}}}(2 i t)\,,
\label{defDm}
\eea
\begin{center}
\begin{figure}[t]
\hspace{2cm}
\includegraphics[scale=.55]{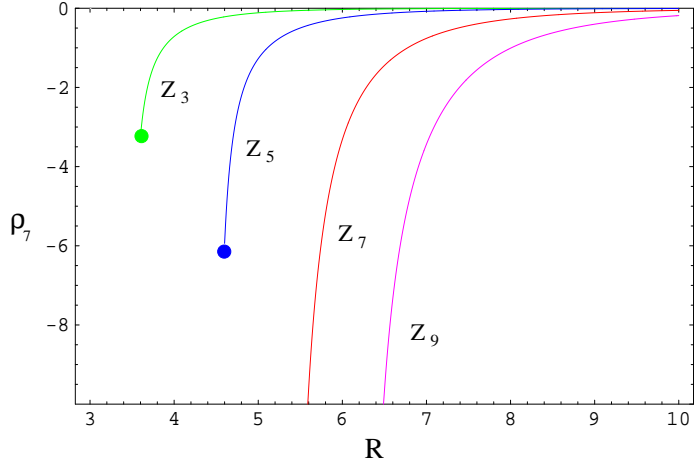}\vspace{-.8cm}
\caption{\small $R$-dependence of the  cosmological constant
for $\Z_N$ orientifolds.
$R$ is the radius of the SS circle, in units of $\al$. The bullets represent the points where
tachyons appear.}
\label{Fig44}
\end{figure}
\end{center}
$\rho_7^{N,K}$ can be in fact nicely combined with $\rho_7^{N,T}$ above, so that
the full closed string contribution reads exactly as (\ref{torozn})
with $d_M(B,F)^k \rightarrow [d_M(B,F)^k+D_M(B,F)^k]/2$.
As expected, the degeneracies of the massless states again agree with the field theory
expectations. This can be easily checked by noting that from (\ref{defDm}) one gets
$D_0(B)^2=D_0(B)^4 = 9$, $D_0(F)^2=D_0(F)^4 =0$ and that
the RR four and zero-form, the NSNS $B$ field and half of the fermions are now
projected out, leaving the decomposition of the remaining states as above.
The open string contribution $\rho_7^{N,(O)}$ is easily computed:
\bea
\rho_7^{N,(O)}=- 2^{-7/2}\int_0^\infty\!\!\frac{dt}{2t^{9/2}}\hspace{-15pt}
\sum_{M\in\mathbb{N},\,n\in\mathbb{Z},\,k=1}^{N-1}\hspace{-12pt}\!
\left[\left(\rm{Tr}\gamma_k\right)^2-(-1)^M\rm{Tr}\gamma_{2k}\right]
e^{2\pi i n \frac{k}{N}}
e^{-2\pi t\alpha^\prime\frac{n^2}{R^2}}D_M^k e^{-2\pi t M},
\label{openzn}
\eea
where the coefficients $D_N^k$ are the same as those defined in (\ref{defDm}) and
$\gamma_k = (\gamma_g)^k$.
Eq.(\ref{openzn}) is not manifestly in the form (\ref{E0-FT3}) because the
embedding of the SS breaking in the gauge sector through the twist matrices $\gamma_g$
requires a little bit of algebra. However, we do not need to work out these details,
since the vacuum energy depends only on the trace of $\gamma_g$.
The integration in $t$ is by now standard and leads to a power-like
behavior in $R$ for the $M=0$ terms and a sum over modified Bessel functions for
$M\neq 0$.

The form of $\rho_7^N(R)$ for various $\Z_N$ orientifold models is reported in figure 3
for a proper choice of $\gamma_g$ that maximizes $\rho_7^N(R)$.
As can be seen from the figure, the qualitative structure of $\rho_7^N$ is not modified
by the orientifold projection. In particular, we still find that $\rho_7^N(R)>\rho_7^M(R)$ for $N<M$.
For different choices of $\gamma_g$, $\rho_7^N$ has always the same form as in figure 3,
but the above ordering of $\rho_7^N$ can be lost.

The vacuum energy density $\rho_7^N(R)$ receives a non-vanishing contribution only
for states that do not propagate along the two twisted directions for both  orbifold
and orientifold models.
This implies that the above computation apply equally well for non-compact $(\mathbb C\times S^1)/\Z_N$
or compact $(T^2 \times S^1)/\Z_N$ models. The energy density is located at the tip of the cone or
at the fixed points of $T^2$, in the two cases. This is expected since these are precisely
the loci where would-be tachyons are localized.

It is interesting to notice that if we plot $\rho_7^N$ as a function of $R/N$, the effective
radius of the SS direction, we get again the behavior as in figures 2 and 3,
but now $\rho_7^N(R/N) < \rho_7^M(R/M)$ for $N<M$, $\forall N,M$ and $R$.
It would be interesting to have some dynamical understanding of
this ordering of the energy densities for this class of models.

The $\Z_2$ orbifold and orientifold models can also be considered. 
In this case space-time is flat with fermions antiperiodic along $R$
and thus we recover exactly the previous 9D model of section 2 or its compactified
version on a $T^2$ torus. 

\section{A four dimensional model}

In this section we study the vacuum energy density of a 4D IIB orientifold model
compactified on $T^6/(\Z_6^\prime\times \Z_2^\prime )$. 
The $\Z_6^\prime$ group is generated by the element $\theta$, acting
as SUSY rotations of angles $2\pi v_i^\theta$ in the three tori $T_i^2\,\, (i=1,2,3)$,
with  $v_i^\theta=1/6(1,2,-3)$ \cite{iba}.  
The complex structure of the first and second torus is fixed by the 
$\Z_6^\prime$ action, whereas the third torus can be taken rectangular,
since the action is a $\Z_2$ reflection.
We define $P_{1,2}$ such that the volume of the $i$-th
torus equals to $4 \pi^2 P_{i}^2/[3(1+\delta_{i,1})]$,
and $R$ and $S$ to be the radii of the two circles 
of $T^2_3$. The $\Z_2^\prime$ group is generated by $\beta = \sigma (-)^F$,
the product of a translation $\sigma$ of length $\pi R$ along the Scherk-Schwarz direction, 
the circle with radius $R$, and $(-)^F$, where $F$ is the 4D space-time fermion number.
The model has $D9$, $D5$ and $\bar D5$-branes. All $D5$-branes are located at the same 
point in space-time and at $y=0$ and $y=\pi R$ along the SS direction.
Similarly, all $\bar D5$-branes are at the same point in space-time and
at $y=\pi R/2$ and $y=3\pi R/2$ along the SS direction.

We  compute the cosmological constant $\rho_4$ as a function of the SS radius and the other moduli. 
Contrary to the previous models, $\rho_4$  gets a non-vanishing
contribution both from states with longitudinal and transverse SS SUSY breaking.
For simplicity  we do not include continuous Wilson lines. A detailed
construction of the model is presented in \cite{adds,sst}.

\subsection{Longitudinal SS breaking contribution}

There are two main longitudinal contributions coming from the torus amplitude
and the open sector of $D9$-branes. 

\paragraph{Closed string contribution.}

Due to the unfolding technique, the longitudinal closed string contribution to the energy 
density, denoted by $\rho_4^{(C,\, l)}$, can be written as 
\be
\rho_4^{(C,\,l)}=-\! \int_0^\infty\!\!\frac{dt}{2 t^3} \sum_i\sum_{j=U,T}
d_i^{(C,j)} e^{-\pi t \al {m_i^2}^{(C,j)}} \sum_{n\in\mathbb Z}
[e^{-\pi t\al{(2n)^2\over R^2}}-e^{-\pi t\al{(2n+1)^2\over R^2}}]\,,
\label{RHO4DCL}
\ee
where $U$ and $T$ stand for the untwisted and $\theta^2/\theta^4$-twisted string sectors---the
only sectors with a non-vanishing contribution---and the index $i$ 
includes the sum over the string, KK and winding modes over the non-SS
directions with the level matching conditions imposed by means of the $\tau_1$ integration\footnote{This 
equals $\bar N=N+\sum_{a=1}^5 m_a n_a$ in the untwisted sector
and $N=\bar N+w_1 n_1$ in the $\theta^2/\theta^4$ twisted ones.}.

The masses of the  $i$-th level for the untwisted ($j=U$) and twisted ($j=T$) sectors are given 
by the following expression:
\bea
{m_i^2}^{(C,j)}\a=\a\frac{2(N+\bar N)}{\al}+\frac{{\al} n_1^2}{S^2}+\frac{w_1^2 S^2}{\al}+
\delta_{j,U}\sum_{a=1}^2\left[\frac{2\al N_a^2}{\sqrt{3}P_a^2}+\frac{\sqrt{3}W_a^2 P_a^2}{2\al} \right],
\eea
where we have defined the combinations of KK and winding modes $N_a^2$ and $W_a^2$ as
\bea
N_a^2\a=\a n_{2a}^2+n_{2a+1}^2+n_{2a}n_{2a+1}\,, \nn\\
W_a^2\a=\a w_{2a}^2+w_{2a+1}^2-w_{2a}w_{2a+1}.
\label{torus4d}
\eea
The degeneracy of the $i$-th state is given by $d_i^{(C,j)}=d^j_N d^j_{\bar N}/24$,
with
\bea
\label{coz6}
\sum_{N,\bar N\in\mathbb N} d^j_N\, d^j_{\bar N} q^N
{\bar q}^{\bar N}=\left\{
\begin{array}{lc}
{|\theta_2^4|^2  | \eta |^{-24}}, &\,\,\,\, j=U\,,
\\
9{|\theta_2^2|^2 | \eta |^{-12}} \left|
{\theta\left[{ 1/2 \atop 1/3}\right] 
\theta\left[{ 1/2 \atop 5/6}\right]^{-1}}
\right|^4, & \,\,\,\,j=T\,.
\end{array}\right.
\eea
Once again, eq.(\ref{RHO4DCL}) is of the same form as (\ref{E0-FT2}).

\paragraph{Open string contribution.}

The longitudinal open string contribution to the energy density, denoted by $\rho_4^{(O,\,l)}$, is
\be
\rho_4^{(O,\,l)}=-\!\int_0^\infty\!\! \frac{dt}{4 (2t)^3} \sum_i \sum_{j=U,T}
d_i^{(O,j)} e^{-2\pi t \al {m_i^2}^{(O,j)}}
\sum_{n\in\mathbb Z} [e^{-\pi t\al{(2n)^2\over R^2}}-e^{-\pi t\al{(2n+1)^2\over R^2}}]\,,
\ee
where
\bea
{m_i^2}^{(O,j)}\a=\a\frac{N}{\al}+\frac{{\al} n_1^2}{S^2}+
\delta_{j,U}\sum_{a=1}^2\frac{2 \al N_a^2}{\sqrt{3} P_a^2},
\eea
with $N_a^2$ as in (\ref{torus4d}) and:
\bea
d_i^{(O,U)}\a=\a \left[\left({\rm Tr}\,\gamma_\beta\right)^2-32 (-1)^N\right]\frac{d_N^U}{6} \,, \\
d_i^{(O,T)}\a =\a
\left[\left({\rm Tr}\,\gamma_\theta^2\gamma_\beta\right)^2-
\left({\rm Tr}\,\gamma_\theta^4\right)(-1)^N\right]
\frac{d_N^T}{6}\,,
\eea
with $\gamma_\theta$ and $\gamma_\beta$ the twist matrices associated respectively
to the SUSY and non-SUSY twists $\theta$ and $\beta$, and $d_N^{U/T}$ defined as in 
(\ref{coz6}).

The untwisted contribution is essentially the same as the open contribution (\ref{apertaIB}) of
the 9D model,  the only difference being an overall factor and the
Kaluza-Klein lattices from the compactified directions.
As in the 9D case, $\gamma_\beta$ is unconstrained (provided that $\gamma_\beta^2=\pm I$).
The value of ${\rm Tr}\,\gamma_\theta^4$ is fixed to be $8$ by tadpole cancellation.

\subsection{Transverse SS breaking contribution}

This contribution arises only when considering $D5$, ${\bar D}5$-branes, as well as
$O5$ and $\bar O5$-planes.
As shown in section 2, these terms can be written as a sum of propagators
of closed string states with mass $m_i$, propagating
along the number of compact dimensions that are orthogonal to the brane.

Tadpole cancellations are automatically encoded when we add up
the Klein bottle, annulus and M\"obius strip contributions in the closed string channel.
Notice that the cancellation is obtained for all the KK modes
along the SS direction due to the choice of the
brane positions, ensuring the local tadpole cancellation in that dimension.

In terms of closed string states, the full amplitude is a sum of two
contributions, one coming from the untwisted and one from the $\theta^2/\theta^4$ twisted string sector.

\paragraph{Untwisted string contribution}

This is given by the sum of the relevant Klein bottle, annulus and M\"obius
strip amplitudes. 
Denoting by $\rho_4^{(U,t)}$ the contribution to the vacuum energy of this sector, we get:
\bea
\label{open4d}
\rho_4^{(U,t)}=-\frac{8 P_2^2 \pi^2}{3\al } \sum_{i,\hat w} d_i^{U} \left\{
8 G^{(i)}_4\left[\Delta x(\hat w)^{(O,U)}\right]+\left[1-2(-1)^N \right]G^{(i)}_4
\left[\Delta x(\hat w)^{(C,U)}\right]\right\}\,,
\eea
where $G_d^{(i)}(x)$ is the $d$-dimensional propagator of a scalar particle of mass ($a=O/C$):
\bea
m_i^{2 (a,\,U)} \a = \a \frac{4N}{\al}+ (1+3\delta_{a,C}) 
\frac{2 P_2^2 W_2^2}{\sqrt{3}} \,, 
\label{m2aU}
\eea
and 
\bea
\left(\Delta x(\hat w)^{(a,U)}\right)^2= \frac{\pi^2 R^2}{\al}\left(w+\frac{1}{2}\right)^2+
(1+3\delta_{a,O})\pi^2 \bigg[\frac{S^2}{\al}n^2 + 
\frac{2 P_1^2 W_1^2}{\sqrt{3}} 
\bigg]\,.
\label{deltaxU}
\eea
\begin{center}
\begin{figure}[t]
\hspace{2cm}
\includegraphics[scale=.55]{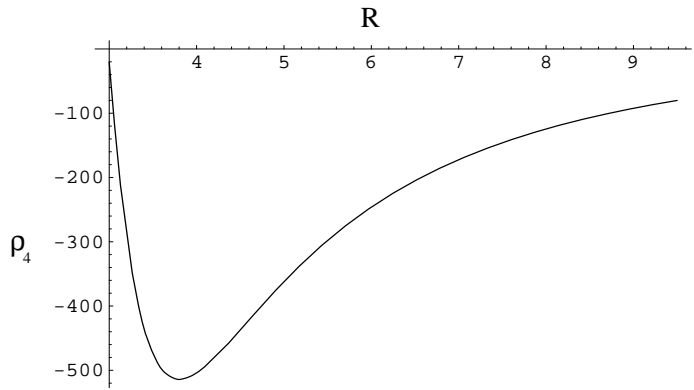}\vspace{-.5cm}
\caption{\small $R$-dependence of the  cosmological constant
for the 4-dimensional model. $R$ refers to the radius of
the Scherk-Schwarz dimension and is in unit of $\al$. The other
moduli are fixed.}
\label{2d}
\end{figure}
\end{center}
In eq.(\ref{open4d}), the three terms in curly brackets are given by the annulus, Klein bottle and
M\"obius strip surfaces, respectively.
Notice that the two labels $O$ and $C$ are needed because the winding and KK modes exchanged
between two $D$-branes (annulus contribution) are not equal to those exchanged between
$O$-planes (Klein bottle) or an $O$-plane and a $D$-brane (M\"obius strip).
More precisely, $O$-planes couple only to even winding modes 
along the longitudinal directions (second torus) and this explains the factor $(1+3\delta_{a,C})$
in (\ref{m2aU}). Similarly, with our choice of $D/\bar D5$-brane positions, closed strings
exchanged between two branes have an integer winding mode along the first torus and the $S$ direction,
whereas half-windings appear between $O$-planes. This explains the factor $(1+3\delta_{a,O})$
in (\ref{deltaxU}).
The degeneracy is given by $d_i^{U}=d_N^U$, with $d_N^U$ defined  in  (\ref{coz6}).

\paragraph{Twisted string contribution}

As before, this is given by the sum of the Klein bottle, annulus and M\"obius
strip amplitudes. It is given by
\bea
\label{open4dtw}
\rho_4^{(T,t)}=-\frac{4 \pi}{3} \sum_{i,\hat w} d_i^{T} \left\{
2 G^{(i)}_2\left[\Delta x(\hat w)^{(O,T)}\right]+\left[1-2(-1)^N \right]G^{(i)}_2
\left[\Delta x(\hat w)^{(C,T)}\right]\right\}\,,
\eea
where the index $i$ runs over $N$, the string oscillator number,
and hence
\bea
m_i^{2 (a,\,T)} \a = \a \frac{4N}{\al}\,,
\eea
independent of $a$, whereas $w$ includes the winding modes in the second torus:
\bea
\left(\Delta x(\hat w)^{(a,T)}\right)^2= \frac{\pi^2R^2}{\al}\left(w+\frac{1}{2}\right)^2+
(1+3\delta_{a,O})
\frac{\pi^2S^2}{\al}w^{\prime\,2} \,.
\label{deltaxT}
\eea
The degeneracy is given by 
\bea
\sum_i d_i^{T}\, e^{-4 \pi i l}=3\,{\frac{\theta_2^2}{\eta^{6}}}\,
\frac{\theta\left[{ 1/6 \atop 0}\right]}{
\theta\left[{ 1/6 \atop 1/2}\right]}
\frac{\theta\left[{ 1/3 \atop 0}\right]}{
\theta\left[{ 1/3 \atop 1/2}\right]} \,.
\eea
All the considerations performed after (\ref{deltaxU}) apply also here,
in order to understand the form of (\ref{open4dtw}) and (\ref{deltaxT}).

It is interesting to note that the transverse contribution is always negative, due to
the fact that the Klein bottle and annulus amplitudes are negative and dominant 
over the  positive M\"obius strip amplitude.

\begin{center}
\begin{figure}[t]
\hspace{.9cm}
\includegraphics[scale=.7]{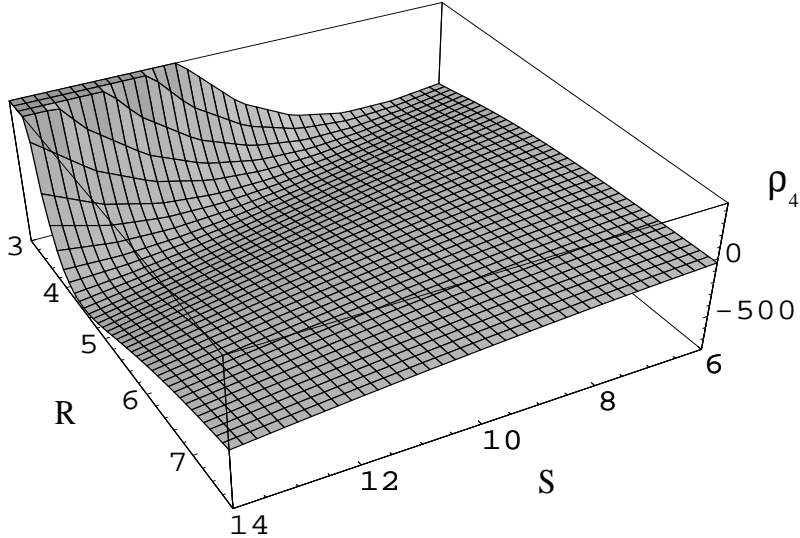}\vspace{-1.5cm}
\caption{\small Dependence of the cosmological constant for the
4-d model on the moduli, expressed in units of $\al$.
R is the Scherk-Schwarz radius and S is the radius of the other
dimensions.}
\label{3d}
\end{figure}
\end{center}

\subsection{Behavior of the cosmological constant}

The full 4D cosmological constant is given by adding together all the contributions, 
coming from both the longitudinal and the transverse SS breaking sectors:
\be
\rho_4 = \rho_4^{(C,\, l)} + \rho_4^{(O,\, l)} + \rho_4^{(U,\, t)} + \rho_4^{(T,\, t)}\,.
\ee
For generic values of ${\rm Tr}\,\gamma_\beta \neq 0$, the longitudinal and
transverse contributions, both negative, result in a monotonic $\rho_4<0$,
leading the system towards the tachyonic instability.
The same instability affects the IIB orbifold model before the $\Omega$-projection,
as can be seen by studying the torus contribution.
Things are more interesting when ${\rm Tr}\, \gamma_\beta = 0 = {\rm Tr}\,\gamma_\theta^2\gamma_\beta$.
In this case, the $D9$ annulus contribution vanishes and the longitudinal
contribution to $\rho_4$ is greater than zero and can thus compensate the
always negative transverse contribution.
This feature is characteristic of a longitudinal breaking when the
SUSY-breaking operator is of order even. In presence of operators of order
odd, as in section 4, it is not possible to take such a choice for
the twist matrices. For this reason, although we have not performed a detailed
analysis, we expect that 4D models with odd SS SUSY twists, such as the model
constructed in \cite{sst}, will unavoidably have $\rho_4<0$ $\forall R$, and
end up in the tachyonic regime.
In the following we thus focus to the case in which both  
$\gamma_\beta$ and $\gamma_\theta^2\gamma_\beta$ are taken to be traceless.

In Fig. \ref{2d} we present the behavior of the one-loop cosmological constant as a function
of the SS radius where the other radii has been fixed to $S=P_1= P_2 \sim 10$ in
units of $\al$. We found a minimum for $\rho_4$ which
is essentially due to a compensation between the longitudinal (positive)
and transverse (negative) monotonic contributions. 

In order to get a better understanding of the fate of this minimum as the other moduli are
varied, we study the behavior of $\rho_4$ as a function of the two parameters
$R$ and $S=P_1=P_2$. The numeric result (figure \ref{3d}) shows that the structure 
of the minimum in $R$ is still preserved as long as $S>8$ but $\rho_4(R,S)$ drives both
moduli to larger values and hence to a decompactification limit.

Other cases similar to the latest one have been considered by studying $\rho_4$ as a function of
two moduli, one of them being the SS circle radius $R$ and the other one the value
of one or more of the extra moduli.
In all  the studies the scenario with a minimum 
in the $R$ direction is preserved, whereas the runaway behavior depends on how 
the extra moduli are fixed. 
As an example, fixing all the moduli in the first and second torus,
$\rho_4(R,S)$ still develops a minimum in $R$ but the behavior 
along $S$ drives the system to smaller radii and thus
towards the tachyonic instability.

\section{Conclusions}

In this paper, the one-loop instabilities of a certain class of IIB orbifold
and orientifold string models have been analyzed. We have shown that typically
in these models the SS direction tends to shrink and to reach the tachyon instability,
as in the early work of \cite{rohm}.
Only for orientifolds with a $\Z_2$ SS twist and with a proper choice of Chan-Paton
twist matrices this situation is modified. In this case
the tachyonic instability is avoided, but the radius increases with a runaway
behavior towards the decompactification limit in which SUSY is restored.
When more geometric moduli are involved, such as in the 4D model of section 5,
the situation is more interesting but less clear. 

Our results are preliminary in various respects. First of all the issue of moduli stabilization
should be considered for all moduli at the same time, both geometrical and not.
This approach is clearly tremendously hard unless some other arguments, like that in \cite{Ginsp},
can predict the behaviour of the vacuum energy density as moduli are 
varied\footnote{See also the recent interesting proposal to stabilize moduli in string theory 
\cite{ags}.}.
In addition, all our results hold at one-loop level only and thus can be spoiled
by higher loop corrections. It has been shown in \cite{add}, for instance, that
higher loop corrections to the vacuum energy density may lead, for large $R$,
to $\log R$ dependencies, in addition to the usual $1/R^n$ terms\footnote{Such 
terms seem to appear also at one-loop level in certain string models 
with uncancelled local tadpoles for $R\rightarrow \infty$ \cite{ablm}.}.
Nevertheless, we think that our results indicate that
the issue of moduli stabilization is somehow more interesting for models admitting
a non-SUSY $\Z_2$ action and for which the corresponding $\rho(R)>0$ for large $R$.
Interestingly, models with a $\Z_2$ non-SUSY action and hence anti-periodic fermions in the bulk
are affected by possible semi-classical instabilities at strong coupling, where
space-time is eaten by a bubble of nothing \cite{Witten}.

In the context of closed string tachyon condensation, it would be interesting 
to see if the jump in the vacuum energy $\rho_7$ when the $\Z_N \rightarrow \Z_{N^\prime}$ 
($N^\prime < N$) transition takes place is exactly accounted for the tachyon condensation.
In this case, assuming that no energy in form of radiation is released in the 
transition (like in the above case of the semi-classical false vacuum decay) and knowing that the final
stage is flat space-time vacuum, one could have some hint on the value of the potential
of some twisted closed string tachyon.

\vskip 25pt
\noindent
{\Large \bf Acknowledgements}
\vskip 10pt

\noindent
We would like to thank M. Blau, D. Perini and C.A. Scrucca for useful discussions.
This work was partially supported by the EC through the RTN network 
``The quantum structure of space-time and the
geometric nature of fundamental interactions'', contract HPRN-CT-2000-00131.
M. B. thanks CONACyT (Mexico).

\appendix

\section*{Appendix}

In the following appendix we briefly discuss the unfolding technique (UT)
used in this paper to calculate the torus contribution to the vacuum
energy density. This technique was first introduced in string theory
in the context of strings at finite temperature \cite{Tan,pol}.
It has also been efficiently used in the context of threshold corrections
to gauge couplings in heterotic string theories (see {\it e.g.} \cite{DKL,MS}, 
and more recently \cite{koko}).

One loop closed string amplitudes on the torus always involve an integral over
the fundamental region $\mathcal F$ which
is given by the complex upper half plane $\tau_2>0$ modded out by the modular group
$PSL(2,\mathbb{Z})$. Analytic integration over this region is
difficult to perform even for zero-point amplitudes, {\it i.e.} in computing the partition function
of a model. The UT gives a systematic procedure that allows one to
unfold $\mathcal F$ into the strip $S$: $\tau_2>0$, $-1/2 \leq \tau_1 \leq 1/2$. In many instances,
such as in the models studied in this paper, such technique is essential to be able
to compute the torus partition function.

The integrand $Z_T$ of a torus partition function is generically given by a sum of terms.
Although $Z_T$ is always modular invariant in any consistent string theory,
each term in the sum generically is not, being mapped to another term 
under the modular group.
It is always possible, however,
to choose a set of representative terms $Z_T^i$ having as an orbit under the action
of the modular group the whole $Z_T$.
We can thus integrate a given set of representatives of the orbit over the unfolding of
$\mathcal F$, namely the strip $S$,
instead of integrating the whole $Z_T$ over $\mathcal F$:
\bea
\int_{\cal F} Z_T=\int_{S/G}\sum_{g\in G} \sum_i g\left[Z_T^i \right]=\int_S \sum_i
Z_T^i \,.
\eea
As an example, we consider the torus amplitude of the 9D model of section 3,
and show in detail how the UT allows to rewrite eq.(\ref{rhod9c}) as (\ref{lambdato}).
As can be easily shown, each of the three terms in curly brackets in (\ref{rhod9c})
is not modular invariant.
The first term is invariant under $\Gamma_0[2]/\Z_2$ which is given by the set of 
unimodular transformations
\bea
\left(\begin{array}{cc}
a&b\\
c&d\\
\end{array}\right)\nn
\eea
with $c$ even and $d$ odd, where $\Z_2$ acts as $a,b,c,d\rightarrow -a,-b,-c,-d$.
The two remaining terms can be reached from the second one by the action of $S$ and $ST$,
where $S(\tau \rightarrow -1/\tau)$ and $T(\tau \rightarrow \tau + 1)$ are the usual
$SL(2,\mathbb Z)$ generators. Moreover, $PSL(2,\mathbb Z)=\{I,S,ST\}\times\Gamma_0[2]/\Z_2$.
The lattice contribution of the first term in (\ref{rhod9c})  can be written  as:
\bea
\frac{\sqrt{\al \tau_2}}{R}\Lambda_{n,w}(-)^n &=&\sum_{n,w\,\in\,\mathbb{Z}} e^{-\frac{\pi R^2}{\al\tau_2}
\left|n+1/2+w\tau\right|^2}=\hspace{-10pt}
\sum_{n\,\in\,\mathbb{Z}\,\mathrm{odd},\,\,w\in\,
\mathbb{Z}\,\mathrm{even}}\hspace{-20pt}
 e^{-\frac{\pi R^2}{4\al\tau_2}\left|n+w\tau\right|^2}\nn\\
&=&
\sum_{p\,\in\,\mathbb{Z}\,\mathrm{odd}}
\sum_{\scriptsize 
\begin{array}{c}
c\in\mathbb{N}\,\mathrm{odd},\\
d\in\mathbb{Z}\,\mathrm{even}
\end{array}}\hspace{-10pt}
 e^{-\frac{\pi R^2}{4\tau_2\al}\,p^2\,\left|c+d\tau\right|^2},
\eea
where we have performed a Poisson resummation and then the change of
variable  $n=pc$,
$w=pd$ with  $c\ge 0$ and  $p$ the minimum common divisor between $n$
and $w$.

By applying  $\tilde\Gamma_0[2]=\Gamma_0[2]/(\Z_2\times T)$ to 
$\sum \exp(-\frac{R^2}{4\al\tau_2}\,p^2)$, we obtain
$\sum \exp(-\frac{R^2}{4\al\tau_2}\,p^2\,\left|c+d\tau\right|^2)$.  Therefore (\ref{rhod9c}) can finally
be written as (\ref{lambdato}):
\bea
\rho_9^{(C)}&=&-R\int_{\cal F}\frac{d^2\tau}{4\tau_2^6}
(1+S+ST)\sum_{g\in \tilde \Gamma_0[2]} g\Bigg[
\frac{\left|\theta_2^4\right|^2}{\left|\eta\right|^{24}}
\sum_{p\in\mathbb{Z}}\frac{1-(-)^p}2 e^{-p^2\frac{\pi R^2}{4\al\tau_2}}\Bigg] \nn\\
&=&-\frac{R}{\sqrt{\alpha^\prime}}\int_S\frac{d^2\tau}{4\tau_2^6}
\frac{\left|\theta_2^4\right|^2}{\left|\eta\right|^{24}}
\sum_{p\in\mathbb{Z}}\frac{1-(-)^p}2
e^{-p^2\frac{\pi R^2}{4 \al\tau_2}}\,,
\eea
where $S$ is the strip  $\tau_2>0$, $-1/2\leq\tau_1\leq 1/2$, the unfolded region of $\mathcal F$.

\end{document}